\documentclass[twocolumn,showpacs,preprintnumbers,amsmath,amssymb]{revtex4}

\usepackage{graphicx}
\usepackage{dcolumn}
\usepackage{bm}

\begin{document}



\title{Observation of Distinct Electron-Phonon Couplings in Gated Bilayer Graphene}

\author{L.~M.~Malard, D. C. Elias, E. S. Alves and M. A. Pimenta}
\address{Departamento de F\'{\i}sica, Universidade Federal de Minas
Gerais, 30123-970, Belo Horizonte, Brazil}

\date{\today}

\begin{abstract}

A Raman study of a back gated bilayer graphene sample is presented.
The changes in the Fermi level induced by charge transfer splits the
Raman G-band, hardening its higher component and softening the lower
one. These two components are associated with the symmetric (S) and
anti-symmetric vibration (AS) of the atoms in the two layers, the
later one becoming Raman active due to inversion symmetry breaking.
The phonon hardening and softening are explained by considering the
selective coupling of the S and AS phonons with interband and
intraband electron-hole pairs.

\end{abstract}

\pacs{63.20.Kd, 78.30.Na, 81.05.Uw}
\maketitle

The interaction of electrons and phonons is a fundamental issue for
understanding the physics of graphene, resulting in the
renormalization of phonon energy and the break-down of the adiabatic
(Born-Oppenheimer) approximation \cite{piscanec04,
andomonolayer,lazzeri06,pisana07}. Although many new physical
insights about the electron-phonon interaction were found in
monolayer graphene \cite{andomonolayer,lazzeri06,pisana07,
yan07,das08}, the bilayer graphene is a unique system to study
distinct couplings between electrons and phonons that have different
symmetries \cite{ando07}. In this paper we show that the application
of a gate voltage in bilayer graphene splits the symmetric and
anti-symmetric optical phonon components, confirming a recent
theoretical prediction for the distinct interactions of these
phonons with intraband and interband electron-hole pairs
\cite{ando07}. This result is specially relevant since the
development of bilayer graphene devices with tunable gap depends on
the detailed understanding of the interaction between electrons and
phonons in this material \cite{castronetoreview, castro07,
vandersypen07}.

The zone center $E_{2g}$ phonon mode of monolayer graphene, which
gives rise to the Raman G-band around 1580 cm$^{-1}$, exhibits a
very strong coupling with electron-hole pairs. This interaction
renormalizes the phonon energy $\hbar \omega_0$, giving rise to the
so-called Kohn anomaly at the zone center of the phonon dispersion
\cite{piscanec04}. Moreover, theoretical models have predicted an
interesting behavior for the phonon energy when the Fermi level
$\varepsilon_F$ is changed by varying the electron or hole
concentration: the phonon energy softens logarithmically for values
of the chemical potential smaller than half of the phonon energy,
and it hardens otherwise. This hardening is ascribed to the
suppression of the interaction between phonons and electron-hole
pairs for $|\Delta\varepsilon_F|$ $>$ $\hbar\omega_0$/2. Besides
this, an increase in either electron of hole density increases the
phonon lifetime \cite{pisana07,andomonolayer,lazzeri06,yan07,das08}
due to the inhibition of the process of phonon decay into
electron-hole pairs, thus reducing the G-band linewidth. Recent
Raman studies of monolayer graphene have confirmed the theoretical
prediction concerning the hardening and narrowing of the the G-band,
by doping the sample with electrons or holes \cite{pisana07, das08,
yan07}.

In the case of bilayer graphene, the $E_{2g}$ phonon mode splits
into two components, associated with the symmetric (S) and
anti-symmetric (AS) displacements of the atoms in the two layers.
Moreover, due to the splitting of the $\pi$ and $\pi^*$ bands in
this material, phonons can couple with electron-hole pairs produced
in interband or intraband transitions. T. Ando \cite{ando07}
calculated recently the self-energy of the S and AS phonons for
varying Fermi energies and predicted the hardening and softening of
the S and AS phonons, respectively, induced by electron or hole
doping. In this work we present experimental results that confirm
the theoretical prediction by T. Ando \cite{ando07}, thus showing
that the G-band of bilayer graphene has indeed two components which
exhibit opposite dependence as the Fermi level is tuned.

\begin{figure}
\includegraphics [scale=0.45]{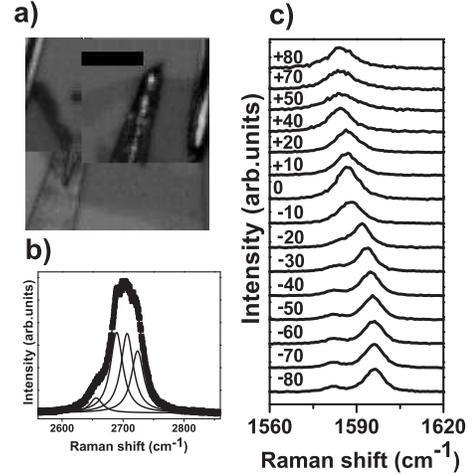}
\caption{\label{Fig1} (a) Optical microscope image of gated graphene
bilayer sample (the black scale bar corresponds to 15 $\mu$m). (b)
Raman spectrum of the G$^{\prime}$ band of bilayer graphene. (c)
Raman spectrum of the G band for different values of the applied
gate voltage.}
\end{figure}

Figure \ref{Fig1}(a) shows an optical microscope image of the
bilayer graphene device obtained by exfoliating bulk graphite on a
300 nm thickness silicon oxide layer on the top of a heavily p doped
silicon substrate. An electrical contact was made by soldering an
indium micro wire directly on the top of the flake \cite{zettl2007}.
Charge carries were induced in the sample by applying a gate voltage
V$_{g}$ to the Si substrate with respect to the graphene contact.
The Raman measurements were carried out at room temperature using a
triple Dilor XY spectrometer with a resolution smaller than 1
cm$^{-1}$, an 80x objective with spot size of $\sim$ 1 $\mu$m and
the 2.41 eV laser excitation with 1 mW power.

The characterization of the number of layers in our graphene sample
was determined by analyzing the shape of the G$^{\prime}$ band
around $\sim$ 2700 cm$^{-1}$ which is shown in Fig. \ref{Fig1}(b)
(this band is also called as D* or 2D by other authors). The
G$^{\prime}$ band of our sample is composed of four peaks and
exhibits a typical shape of a bilayer graphene
\cite{ferrari07,gupta06,graf07,malard07}.

Figure \ref{Fig1}(c) shows the Raman spectra taken at different
applied gate voltages. We observe in this figure that both the
position and the shape of the G-band depend on the applied gate
voltage, in agreement with previous Raman studies of gated bilayer
graphene \cite{yan08, das08b}. However, as our experiments were done
at room temperature, we were not able to observe the initial
softening of the G-band for $|\Delta \varepsilon_F| < \hbar
\omega_G$/2, as observed at low temperatures by Yan et al.
\cite{yan08}.

\begin{figure}
\includegraphics [scale=0.45]{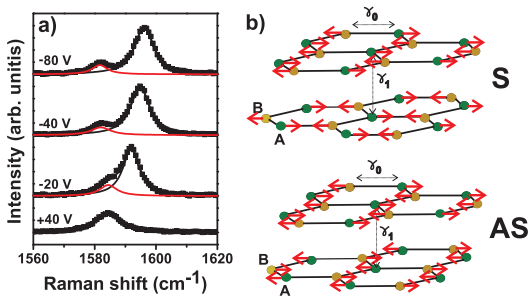}
\caption{\label{Fig2} (color online)(a) Raman G-band of the bilayer
graphene for -80 V, -40 V, -20 V and +40 V gate voltages . Two
Lorentzian curves are needed to fit the G-band for -80 V, -40 V and
-20 V. (b) Displacement of the atoms for the symmetric and
anti-symmetric highest energy phonon modes in the $\Gamma$ point of
bilayer graphene.}
\end{figure}

Figure \ref{Fig2}(a) shows the fittings of the Raman G-band of the
bilayer graphene device for different values of applied gate
voltage. The G-bands observed in the spectra taken at V$_{g}$ equals
to -80 V, -40 V and -20 V clearly exhibit two peaks. All spectra
taken at positive values of gate voltage could be fitted by a single
Lorentzian.

Figures \ref{Fig3}(a) and \ref{Fig3}(b) show the relative shift
(with respect to G-band position in the + 50 V spectrum) and the
full width at half maximum (FWHM) of the Lorentzians that fit the
G-band as a function of the applied gate voltage, respectively.
Notice that the frequency of the high energy component increases
with decreasing values of applied gate voltage, whereas the opposite
behavior is observed for the low energy component. Figure
\ref{Fig3}(b) shows that the FWHM exhibits a maximum for V$_{g}= +
50$ V spectrum, and it clearly decreases with decreasing values of
applied gate voltages.

Let us discuss the origin of the two peaks that compose the G-band
shown in Fig. \ref{Fig2}(a). The bilayer graphene is formed by two
graphene layers in the AB Bernal stacking, where the point group at
$\Gamma$ point is D$_{3d}$. The phonon branch associated with the
$E_{2g}$ mode of monolayer graphene gives rise to two branches for
bilayer graphene, one symmetric (S) and other anti-symmetric (AS)
components (in-phase and out-of-phase displacements of the atoms in
the two layers), which are represented at Fig. \ref{Fig2}(b). At the
center of the Brillouin zone ($\Gamma$ point), the symmetric and
anti-symmetric vibrations belong to the two doubly degenerated
representations E$_{g}$ and E$_{u}$, respectively. The
anti-symmetric E$_{u}$ mode is not Raman active since the D$_{3d}$
group is centro-symmetric. However, if the inversion symmetry
operation of the bilayer graphene is broken, the system is described
by the C$_{3v}$ point group at the $\Gamma$ point.  In this case,
both the symmetric and the anti-symmetric modes belong to the E
representation, and are Raman active.

\begingroup
\begin{figure}
\centering
\includegraphics [scale=0.5]{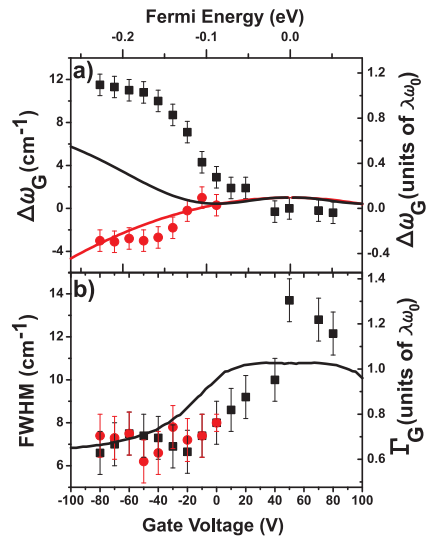}
\caption{\label{Fig3} (color online)(a) The black squares and red
circles correspond to the relative shift ($\Delta\omega_{G}$) of the
symmetric and anti-symmetric components of the G band, respectively,
with respect to the position in the +50 V spectrum, as a function of
the applied gate voltage (bottom scale).  The solid curves
corresponds to the theoretical prediction for the relative G-band
shift \cite{ando07} in units of $\lambda \omega_0$ as a function of
the Fermi energy (top scale). (b)  The black squares and red circles
correspond to the FWHM of the symmetric and anti-symmetric
components of the G-band, respectively, and the solid curve
corresponds to the theoretical prediction for the symmetric phonon
mode \cite{ando07}.}
\end{figure}
\endgroup

The inversion symmetry breaking can be due to the different
materials that the top and bottom graphene layers of our device are
exposed to and/or to a non-homogeneous doping of the top and bottom
layers. The non-equivalence between the top and bottom layers
decreases the symmetry, making the anti-symmetric mode active in the
Raman spectra, explaining thus the two components of the G-band. The
energy separation between the S and AS components is negligible when
the Fermi level is close to the Dirac point, and this can be seen in
the spectrum taken at + 40 gate voltage in Fig. \ref{Fig1}(b).
However, the doping with electrons or holes induces a measurable
energy splitting for those components, as shown in the spectra
between -20 V to - 80 V in Fig. \ref{Fig1}(b).

One could argue that the observed splitting of the G band is related
to symmetric mode coming from the top and bottom layers, in the case
of a non-homogeneous doping of the two layers. However, this
interpretation has serious drawbacks in the present case. Notice
that the low energy component of the G-band is about five times less
intense than the higher energy one. Considering that the
transmitance of visible light in monolayer graphene is about
97.7$\%$  \cite{nair08}, the contributions of symmetric mode from
the upper and bottom layers for the Raman scattering are expected to
be almost the same. Furthermore, our results show that the energy of
the low energy component decreases when the Fermi level is changed.
This result would not be expected if one associates the lower energy
Raman peak with the symmetric mode of the bilayer graphene bottom
layer. We stress the fact that the splitting of the G-band was not
reported in the two recent studies of gated bilayer graphene devices
\cite{yan08, das08b}, despite the fact that the G-band observed in
reference \cite{yan08} is clearly asymmetric. Possibly, our
observation of this splitting is due to the characteristic of our
device which leads to the inversion symmetry breaking and allows the
observation of the non-active Raman AS vibration.

\begin{figure}[h]
\includegraphics [scale=0.45]{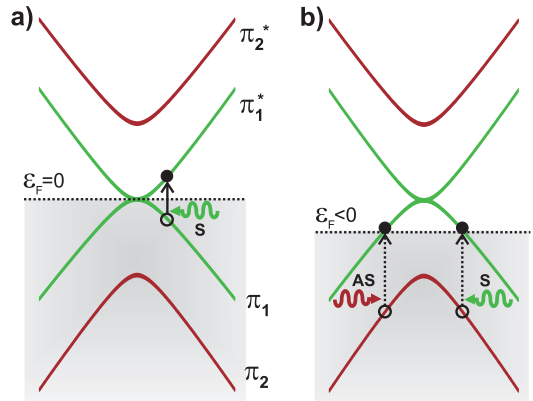}
\caption{\label{Fig4} (color online) Parabolic band structure of
bilayer graphene near the K point. The vertical arrows illustrates
the possible transitions induced by symmetric (green) and
antisymmetric (red) $\textbf{q}$=0 phonons for (a) interband
electron-hole pairs creation at ($\varepsilon_{F}$=0) and (b)
intraband electron-hole pairs creation ($\varepsilon_{F}<$0). The
gap opening is not considered in this diagram.}
\end{figure}

In order to discuss the phonon renormalization effect in bilayer
graphene, we must consider the selection rules for the interaction
of the symmetric and anti-symmetric phonons with the interbands or
intrabands electron-hole pairs. The band structure of bilayer
graphene near the K point shown in Figure \ref{Fig4} consists of
four parabolic bands, two of them touch each other at the K point,
and the other two bands separated by 2$\gamma_{1}$, where
$\gamma_{1}\sim 0.35$ eV. The electron-phonon interaction in bilayer
graphene is described by a 2x2 matrix for each phonon symmetry,
where each matrix element gives the contribution of electron-hole
pairs involving different electronic sub-bands \cite{ando07}. For
the symmetric phonon mode, all matrix elements are different from
zero, and this phonon can interact with both interband or intraband
electron-hole pairs, as shown in Figure \ref{Fig4}(a), giving rise
to the phonon energy renormalization (Kohn anomaly). However, for
the anti-symmetric phonon mode, the diagonal terms of the matrix are
null, showing that there is no coupling between AS phonons and
interband electron-hole pairs. Therefore, no Kohn anomaly is
expected for the antisymmetric phonon mode when the Fermi level is
at the Dirac point.

However, if the Fermi energy is changed [for instance, $
\varepsilon_{F}<$ 0 as shown in Fig. \ref{Fig4}(b)], intraband
electron-hole pairs can be produced by phonons. Now, the
anti-symmetric phonons have also their energies renormalized, giving
rise to the Kohn anomaly. Notice that as the energy separation
between the $\pi_{1}$ and $\pi_{2}$ bands ($\gamma_{1}\sim$ 0.35 eV)
is larger than the G band energy ($\sim$ 0.2 eV), the electron-hole
pairs creation by those phonons is a virtual process. Therefore, we
expect the dependence of the shift of the anti-symmetric mode with
Fermi energy to be smoother than that of the symmetric mode. A
similar result has been recently observed in gated semiconducting
carbon nanotubes \cite{avouris2007} which have the band gap larger
than the phonon energy.

In order to explain our experimental results, we have also plotted
in Figs. Fig. \ref{Fig3} (a) and (b) frequency shift and broadening
of the G band calculated by T. Ando \cite{ando07} as
$\Delta\omega_G=Re[\Pi^{(\pm)}(\omega)]$ and
$\Gamma_G=-Im[\Pi^{(\pm)}(\omega)]$ respectively, where $\Pi$ is the
phonon self energy given of the symmetric (+) or anti-symmetric (-)
components, given by :

\begingroup
\begin{eqnarray}\label{eq1}
 \Pi^{(\pm)}(\omega)=-\lambda
\int^{\infty}_{0}&\gamma^{2} kdk\sum_{j,j^{\prime}}
\sum_{s,s^{\prime}}\Phi^{(\pm)}_{jj^{\prime}(k)}
\nonumber\\
&\times\frac{[f(\varepsilon_{sjk})-f(\varepsilon_{s^{\prime}j^{\prime}k^{\prime}})]
(\varepsilon_{sjk}-\varepsilon_{s^{\prime}j^{\prime}k^{\prime}})}{(\hbar\omega+i\delta)^2-
(\varepsilon_{sjk}-\varepsilon_{s^{\prime}j^{\prime}k^{\prime}})^2}.
\end{eqnarray}
\endgroup
The functions $\varepsilon_{sjk}$ are the electronic band
dispersions ($k$ is the modulus of the wavevector), $s$ = +1 and -1
denotes the conduction and valence bands, respectively, and $j$ = 1,
2 specifies the two bands within the valence and conduction bands.
The function $f(\varepsilon_{sjk})$ is the Fermi distribution and
the matrix elements $\Phi^{\pm}_{jj^{\prime}}(k) $ gives the
relative contribution of the S (+) and AS (-) phonons for the
interband and intraband electron-hole pairs formed in transitions
\cite{ando07}. The damping parameter $\delta$ describes the charge
inhomogeneity in the sample and washes out the logarithmic
singularity present at $\varepsilon_{F}=\hbar\omega_{0}/2$. We have
considered the value of $\delta$ = 0.1 eV in the theoretical curves
depicted in Fig. \ref{Fig3}.

Theoretical curves $\Delta\omega_G$ and $\Gamma_G$ are calculated in
units of $\lambda
 \omega_0$, where $\omega_0$ = 0.196 eV is the energy of the
optical phonon and $\lambda$ is related to the strength of the
electron-phonon coupling and given by \cite{ando07}:
\begin{equation}
\lambda= 0.16\times 10^{-3}({\AA}^2eV^{-2})[\frac{\partial
\gamma_0}{\partial b}]^2 \label{eq2}
\end{equation}

In order to compare the experimental and theoretical results
depicted in Figs. \ref{Fig3} (a) and (b) we need to convert the
experimental horizontal scale (applied gate voltage, in the bottom
axis of Fig. \ref{Fig3}) into the theoretical scale (Fermi energy,
in the upper axis of Fig. \ref{Fig3}), and the experimental vertical
scales (in cm$^{-1}$, left side axes of Fig. \ref{Fig3}) into the
theoretical scales (in units of $\lambda \omega_0$, right side axes
of Fig. \ref{Fig3}).

To match the scales of the experimental (left axis) and calculated
(right axis) values of $\Delta\omega_G$ and $\Gamma_G$, shown in
Fig.~\ref{Fig3}, we have used the value $\partial \gamma_0 /
\partial b$ = 6.4 eV ${\AA}^{-1}$ in Eq.~\ref{eq2}
\cite{castroneto07}. Both theoretical curves $\Delta\omega_G$ and
$\Gamma_G$ are plotted in Fig.~\ref{Fig3} as a function of the Fermi
energy $\varepsilon_{F}$ (top axis), while the experimental values
are plotted as a function of gate voltage (bottom axis). In order to
scale these axes, we have used the fact that both $\varepsilon_{F}$
and V$_{g}$ are related to the electron (or hole) density $n$, which
is assumed to be the same in both layers. The modulus of the Fermi
energy $|\varepsilon_F|$ is related to the electron (or hole)
density in the regime $|\varepsilon_{F}|<\gamma_{1}$ by
\cite{ando07}:
\begin{equation}
|\varepsilon_{F}|=\frac{1}{2}(-\gamma_{1}+\sqrt{4n\pi\gamma^{2}+\gamma_{1}^{2}}),
\end{equation}
where $\gamma = \frac{\sqrt{3}}{2}a\gamma_{0}$, $a = 2.42 {\AA}$ is
the lattice constant, $\gamma_{0}\approx 3 $eV is the in-plane
nearest-neighbor tight-binding parameter and $\gamma_{1} \sim 0.35$
eV is the out-of-plane nearest-neighbor parameter [see Fig.
\ref{Fig2}(b)]. A parallel plate capacitor model gives
$n=7.2\times10^{10}$cm$^{-2}V^{-1}(V_{g}-V_{D})$, where $V_{D}$ is
the gate voltage needed to move the Fermi level to the Dirac point
due to intrinsic doping of the sample. The value of $V_{D}$ = + 50 V
was obtained from the spectrum where the G-band has the largest FWHM
[see Fig. \ref{Fig3}(b)], since the symmetric phonon lifetime is
minimum for $\varepsilon_F$ = 0 \cite{ando07}.

Figure~\ref{Fig3} shows that there is a good qualitative agreement
between our experimental results and the theoretical prediction of
the dependence of the frequency and width of the symmetric and
anti-symmetric optical phonons in bilayer graphene with the Fermi
level. We shall stress that this good agreement is obtained despite
the different approximations that have been done. Notice first that
the theoretical model doesn't take into account the trigonal warping
effect, the angular dependence of the electron-phonon coupling, and
the strength $\partial \gamma_1 /
\partial b$ for the coupling between electrons and AS phonons [see
Fig.~\ref{Fig2}(b)]. Moreover, we have used the simple model of a
parallel capacitor with homogeneous carrier concentration in the two
layers. A more complete model is needed to improve the fitting of
the experimental data.

In summary, we have shown in this work that bilayer graphene is a
unique material where the phonon renormalization tuned by charge
transfer depends strongly on the symmetry of the phonons involved in
the creation of electron-hole pairs. The inversion symmetry breaking
of our bilayer graphene device allows the observation of both the
symmetric and anti-symmetric phonon modes in the Raman G band. The
hardening of the symmetric mode and the softening of the
antisymmetric phonon mode when the sample is doped is explained by
the selection rules associated with the creation of interband or
intrabands electron-hole pairs by symmetric and anti-symmetric
optical phonons, giving an experimental support for the theoretical
predictions  of the optical phonons dependence on charge
concentration in bilayer graphene \cite{ando07}.

We would like to acknowledge the very useful discussions with Profs.
R. Saito, A. H. Castro Neto and K. S. Novoselov. The graphite sample
used to prepare the exfoliated graphene was provided by Nacional de
Grafite (Brazil). This work was supported by Rede Nacional de
Pesquisa em Nanotubos de Carbono - MCT and FAPEMIG. L.M.M. and D.C.E
acknowledges the support from the Brazilian Agency CNPq.

\end{document}